\documentclass[12pt]{article}
\newcommand{\bq}{\begin{equation}}
\newcommand{\eq}{\end{equation}}
\newcommand{\ba}{\begin{eqnarray}}
\newcommand{\ea}{\end{eqnarray}}

\newcommand{\nl }{ \nonumber  }

\newcommand{\p}{\partial}

\newcommand{\h}{\hspace{.5cm}}
\newcommand{\s}{\sigma}

\begin{document}
\begin{flushright}
{\bf hep-th/0008006}
\end{flushright}
\vspace*{1.cm}
{\bf\begin{center}
EXACT STRING SOLUTIONS IN CURVED BACKGROUNDS
\vspace*{0.5cm}\\ P. Bozhilov \footnote {E-mail:
bojilov@thsun1.jinr.ru}, \\ \it Bogoliubov Laboratory of Theoretical
Physics, \\ JINR, 141980 Dubna, Russia
\end{center}}
\vspace*{0.5cm}
We show how the classical string dynamics in $D$-dimensional curved
background can be reduced to the dynamics of a massless particle constrained
on a certain surface whenever there exists at least one Killing vector for
the background metric. Then we obtain a number of sufficient conditions,
which ensure the existence of exact solutions to the equations of motion and
constraints. The results are also relevant to the null string case. Finally,
we illustrate our considerations with an explicit example in four dimensions.
\\

PACS number(s): 04.25.-g, 11.10.Lm, 11.25.-w, 11.27.+d


\vspace*{.5cm}

\section{Introduction}
\hspace{1cm}The string equations of motion and constraints in curved
space-time are highly nonlinear and, {\it in general}, non exactly solvable
\cite{LS953}-\cite{dVS952}.
Different methods have been applied to solve them approximately
\cite{dVS87}, \cite{GSV91}, \cite{dVN92}-\cite{CLS96} or, if possible, exactly
\cite{MR92}-\cite{MO00}.
On the other hand, quite general exact solutions can be found
by using an appropriate ansatz, which exploits the symmetries of the underlying
curved space-time \cite{dVS94}, \cite{dVE93}-\cite{dVE95}.
In most cases, such an ansatz effectively decouples the dependence on the
spatial world-sheet coordinate $\sigma$ \cite{dVE93}-\cite{LS00} or
the dependence on the temporal world-sheet coordinate $\tau$
\cite{LS96,LS98}, \cite{LS951}-\cite{dVE96}.
Then the string equations of motion and constraints reduce to nonlinear
coupled {\it ordinary} differential equations, which are considerably simpler
to handle than the initial ones.

In this letter, we obtain some exact solutions of the classical equations
of motion and constraints for both tensile and null strings in a
$D$-dimensional curved background. This is done by using an
ansatz, which reduces the initial dynamical system to the one depending on
only one affine parameter. This is possible whenever there exists at least
one Killing vector for the background metric. Then we search for sufficient
conditions, which ensure the existence of exact solutions to the equations of
motion and constraints without fixing particular metric. After that, we give
an explicit example in four dimensional Kasner type background. Finally,
we conclude with some comments on the derived results.

\section{Reduction of the dynamics}
\hspace{1cm}To begin with, we write down the bosonic string action in $D$-
dimensional curved space-time $\mathcal{M}_D$ with metric tensor
$g_{MN}\left(X\right)$
\ba\label{as} S&=&\int d^{2}\xi {\cal L},\h
{\cal L}= -\frac{T}{2}\sqrt{-\gamma}\gamma^{mn}
\p_m X^M\p_n X^N g_{MN}\left(X\right),
\\ \nl
\p_m &=&\p/\p\xi^m,\h \xi^m=(\xi^0,\xi^1)=(\tau,\s),\h
m,n=0,1,\h M,N=0,1,...,D-1, \ea
where, as usual, $T$ is the string tension and $\gamma$ is the determinant
of the induced metric.

Here we would like to consider tensile and null (tensionless) strings
on equal footing, so we have to rewrite the action (\ref{as}) in a form
in which the limit $T\to 0$ can be taken. To this end, we set
\ba\nl \gamma^{mn}=
\left(\begin{array}{cc}-1&\lambda^1\\
\lambda^1&-\left(\lambda^1\right)^2 +
\left(2\lambda^0 T\right)^2\end{array}\right)\ea
and obtain
\ba\nl {\cal L}= \frac{1}{4\lambda^0}g_{MN}\left(X\right)
\left(\p_0-\lambda^1\p_1\right) X^M\left(\p_0-\lambda^1\p_1\right) X^N
- \lambda^0 T^2 g_{MN}\left(X\right)\p_1X^M\p_1X^N.\ea
The equations of motion and constraints following from this Lagrangian
density are
\ba\nl \p_0\left[ \frac{1}{2\lambda^0}\left(\p_0-\lambda^1\p_1\right)
X^K\right] - \p_1\left[ \frac{\lambda^1}{2\lambda^0}
\left(\p_0-\lambda^1\p_1\right)X^K\right]\\ \nl + \frac{1}{2\lambda^0}
\Gamma^{K}_{MN}\left(\p_0-\lambda^1\p_1\right) X^M
\left(\p_0-\lambda^1\p_1\right) X^N = \\ \nl 2\lambda^0 T^2
\left[ \p_1^2X^K + \Gamma^{K}_{MN}\p_1X^M\p_1X^N +
\left(\p_1\ln\lambda^0\right)\p_1X^K\right],\\
\label{c1}g_{MN}\left(X\right)
\left(\p_0-\lambda^1\p_1\right) X^M
\left(\p_0-\lambda^1\p_1\right) X^N =
-\left(2\lambda^0 T\right)^2 g_{MN}\left(X\right)\p_1X^M\p_1X^N,\\
\label{c2}g_{MN}\left(X\right)\left(\p_0-\lambda^1\p_1\right) X^M\p_1X^N=0,
\ea
where
\ba\nl\Gamma^{K}_{MN} = \frac{1}{2}g^{KL}\left(\p_Mg_{NL} +
\p_Ng_{ML} - \p_Lg_{MN}\right)\ea
is the connection compatible with the metric $g_{MN}\left(X\right)$.
We will work in the gauge $\lambda^m = constants$ in which the Euler-Lagrange
equations take the form
\ba\label{ele}\left(\p_0-\lambda^1\p_1\right)
\left(\p_0-\lambda^1\p_1\right)X^K + \Gamma^{K}_{MN}
\left(\p_0-\lambda^1\p_1\right) X^M
\left(\p_0-\lambda^1\p_1\right) X^N \\ \nl = \left(2\lambda^0 T\right)^2
\left( \p_1^2X^K + \Gamma^{K}_{MN}\p_1X^M\p_1X^N \right).\ea

Now we are going to show that by introducing an appropriate ansatz, one can
reduce the classical string dynamics to the dynamics of a massless particle
constrained on a certain surface whenever there exists at least one Killing
vector for the background metric. Indeed, let $M=(\mu,a)$,
$\{\mu\}\neq\{\emptyset\}$ and let us suppose that there exist a number of
independent Killing vectors $\eta_{\mu}$. Then in appropriate coordinates
$\eta_{\mu}=\frac{\p}{\p X^{\mu}}$ and the metric does not depend on
$X^{\mu}$. In other words, from now on we will work with the metric
\ba\nl g_{MN}=g_{MN}\left(X^a\right).\ea
On the other hand, we observe that
\ba\nl X^M(\tau,\sigma) = F^{M}_{\pm}[w_{\pm}(\tau,\sigma)],\h
w_{\pm}(\tau,\sigma)=(\lambda^1\pm2\lambda^0 T)\tau + \sigma \ea
are solutions of the equations of motion (\ref{ele}) in arbitrary $D$-
dimensional background $g_{MN}\left(X^K\right)$, depending on $D$
arbitrary functions $F^{M}_{+}$ or $F^{M}_{-}$ (see also \cite{MR92} for
the possible background independent solutions and their properties).
Taking all this into account, we propose the ansatz
\ba\label{ta} X^{\mu}\left(\tau,\sigma\right) &=&
C^{\mu}_{\pm}w_{\pm} + y^{\mu}\left(\tau\right),\h C^{\mu}_{\pm}=constants,
\\ \nl  X^{a}\left(\tau,\sigma\right) &=& y^{a}\left(\tau\right).\ea
Inserting (\ref{ta}) into constraints (\ref{c1}) and (\ref{c2}) one obtains
(the dot is used for $d/d\tau$)
\ba\nl g_{MN}(y^a)\dot{y}^M \dot{y}^N \pm 2\lambda^0 TC^{\mu}_{\pm}
\left[g_{\mu N}(y^a)\dot{y}^N \pm 2\lambda^0 TC^{\nu}_{\pm}
g_{\mu\nu}(y^a)\right] = 0,\\ \nl C^{\mu}_{\pm}
\left[g_{\mu N}(y^a)\dot{y}^N \pm 2\lambda^0 TC^{\nu}_{\pm}
g_{\mu\nu}(y^a)\right] = 0.\ea
Obviously, this system of two constraints is equivalent to the following one
\ba\label{tc1'} g_{MN}(y^a)\dot{y}^M \dot{y}^N = 0,\\
\label{tc2} C^{\mu}_{\pm}
\left[g_{\mu N}(y^a)\dot{y}^N \pm 2\lambda^0 TC^{\nu}_{\pm}
g_{\mu\nu}(y^a)\right] = 0.\ea
Using the ansatz (\ref{ta}) and constraint (\ref{tc2}) one can reduce
the initial Lagrangian to get
\ba\nl L^{red}(\tau)\propto \frac{1}{4\lambda^0}
\left[g_{MN}(y^a)\dot{y}^M\dot{y}^N - 2\left(2\lambda^0 T\right)^2
C^{\mu}_{\pm}C^{\nu}_{\pm}g_{\mu\nu}(y^a)\right].\ea
It is easy to check that the constraint (\ref{tc1'})
can be rewritten as $g^{MN}p_M p_N=0$, where $p_M=\p L^{red}/\p\dot{y}^M$
is the momentum conjugated to $y^M$. All this means that we have obtained
an effective dynamical system describing a massless point particle moving
in a gravity background $g_{MN}(y^a)$ and in a potential
\ba\nl U\propto T^2 C^{\mu}_{\pm}C^{\nu}_{\pm}g_{\mu\nu}(y^a)\ea
on the constraint surface (\ref{tc2}).

Analogous results can be received if one uses the ansatz
\ba\label{sa} X^{\mu}\left(\tau,\sigma\right) =
C^{\mu}_{\pm}w_{\pm} + z^{\mu}\left(\sigma\right),\h
X^{a}\left(\tau,\sigma\right) = z^{a}\left(\sigma\right).\ea
Now putting (\ref{sa}) in (\ref{c1}) and (\ref{c2}) one gets
($'$ is used for $d/d\sigma$)
\ba\nl &&\left[\left(2\lambda^0 T\right)^2 + (\lambda^1)^2\right]
g_{MN}z'^Mz'^N \\ \nl
&&+ 4\lambda^0 T\left[\left(2\lambda^0 T\mp\lambda^1\right)
C^{\mu}_{\pm}g_{\mu N}z'^N
+ 2\lambda^0 TC^{\mu}_{\pm}C^{\nu}_{\pm}
g_{\mu\nu}\right]=0,\\ \nl
&&\lambda^1 g_{MN}z'^Mz'^N + \left[\left(\lambda^1\mp2\lambda^0 T\right)
C^{\mu}_{\pm}g_{\mu N}z'^N \mp 2\lambda^0 TC^{\mu}_{\pm}C^{\nu}_{\pm}
g_{\mu\nu}\right]=0.\ea
These constraints are equivalent to the following ones
\ba\nl  g_{MN}(z^a)z'^Mz'^N =0,\\
\label{sc2'} C^{\mu}_{\pm}\left[g_{\mu N}(z^a)z'^N +
\frac{2\lambda^0 T}{2\lambda^0 T\mp\lambda^1}
C^{\nu}_{\pm}g_{\mu\nu}(z^a)\right]=0.\ea
The corresponding reduced Lagrangian obtained with the help of (\ref{sa}) and
(\ref{sc2'}) is
\ba\nl L^{red}(\sigma)\propto \frac{\left(\lambda^1\right)^2 -
\left(2\lambda^0 T\right)^2}{4\lambda^0}\left[g_{MN}(z^a)z'^Mz'^N -
2\left(\frac{2\lambda^0 T}{2\lambda^0 T\mp\lambda^1}\right)^2
C^{\mu}_{\pm}C^{\nu}_{\pm}g_{\mu\nu}(z^a)\right]\ea
and a similar interpretation can be given as before.

In both cases - the ansatz (\ref{ta}) and the ansatz (\ref{sa}),
the reduced Lagrangians do not depend on $y^\mu$ and $z^\mu$ respectively,
and their conjugated generalized momenta are conserved.

Let us point out that the main difference between tensile and null strings,
from the point of view of the reduced Lagrangians, is the absence of a
potential term for the latter.

Because the consequences of (\ref{ta}) and (\ref{sa}) are similar,
our further considerations will be based on the ansatz (\ref{ta}).

\section{Exact solutions}
\hspace{1cm} To obtain the equations which we are going to consider, we
use the ansatz (\ref{ta}) and rewrite (\ref{ele}) in the form
\ba\label{le} g_{KL}\ddot{y}^L + \Gamma_{K,MN}\dot{y}^M\dot{y}^N \pm
4\lambda^0 TC^{\mu}_{\pm}\Gamma_{K,\mu N}\dot{y}^N = 0.\ea
At first, we set $K=\mu$ in the above equality. It turns out that in this
case the equations (\ref{le}) reduce to
\ba\nl \frac{d}{d\tau}\left[g_{\mu\nu}\dot{y}^{\nu} + g_{\mu a}\dot{y}^a
\pm 2\lambda^0 TC^{\nu}_{\pm}g_{\mu\nu}\right]=0,\ea
i.e we have obtained the following first integrals (constants of the motion)
\ba\label{fi} g_{\mu\nu}\dot{y}^{\nu} + g_{\mu a}\dot{y}^a
\pm 2\lambda^0 TC^{\nu}_{\pm}g_{\mu\nu} = A^{\pm}_{\mu} = constants.\ea
They correspond to the conserved momenta $p_\mu$. From the constraint
(\ref{tc2}) it follows that the right hand side of (\ref{fi}) must satisfy
the condition
\ba\nl C^{\mu}_{\pm}A^{\pm}_{\mu} = 0.\ea

Using (\ref{fi}), the equations (\ref{le}) for $K=a$ and the constraint
(\ref{tc1'}) can be rewritten as
\ba\label{ema} 2\frac{d}{d\tau}\left(h_{ab}\dot{y}^b\right) -
\left(\p_a h_{bc}\right)\dot{y}^b\dot{y}^c + \p_a V = 4\p_{[a}
\left(g_{b]\mu}k^{\mu\nu}A^{\pm}_{\nu}\right)\dot{y}^b\ea
and
\ba\label{c2a} h_{ab}\dot{y}^a\dot{y}^b + V =0,\ea where
\ba\nl h_{ab}\equiv g_{ab} - g_{a\mu}k^{\mu\nu}g_{\nu b},\h\h
V\equiv A^{\pm}_{\mu}A^{\pm}_{\nu}k^{\mu\nu} +
\left(2\lambda^0 T\right)^2C^{\mu}_{\pm}C^{\nu}_{\pm}g_{\mu\nu},\ea
and $k^{\mu\nu}$ is by definition the inverse of $g_{\mu\nu}$:
$k^{\mu\lambda}g_{\lambda\nu}=\delta^{\mu}_{\nu}$. For example, when $g_{MN}$
does not depend on the coordinate $y^q$
\ba\nl h_{ab}= g_{ab} - \frac{g_{aq}g_{qb}}{g_{qq}},\ea
when $g_{MN}$ does not depend on two of the coordinates (say $y^q$ and
$y^s$)
\ba\nl h_{ab}= g_{ab} - \frac{g_{aq}g_{ss}g_{qb} - 2g_{aq}g_{qs}g_{sb}
+ g_{as}g_{qq}g_{sb}}{g_{qq}g_{ss} - g_{qs}^{2}},\ea and so on.

At this stage, we restrict the metric $h_{ab}$ to be a diagonal one, i.e.
\ba\label{mc} g_{ab} = g_{a\mu}k^{\mu\nu}g_{\nu b},
\h\h \mbox{for}\h\h a\ne b.\ea
This allows us to transform further equations (\ref{ema}) and obtain
(there is no summation over $a$)
\ba\label{emat} \frac{d}{d\tau}\left(h_{aa}\dot{y}^a\right)^2 +
\dot{y}^a\p_a\left(h_{aa} V\right)
+ \dot{y}^a\sum_{b\ne a}
\left[\p_a\left(\frac{h_{aa}}{h_{bb}}\right)\left(h_{bb}\dot{y}^b\right)^2
- 4\p_{[a}A^{\pm}_{b]} h_{aa}\dot{y}^b\right] = 0,\ea
where we have introduced the notation
\ba\label{gp} A^{\pm}_{a} \equiv g_{a\mu}k^{\mu\nu}A^{\pm}_{\nu}.\ea
In receiving (\ref{emat}), the constraint (\ref{c2a}) is also used after
taking into account the restriction (\ref{mc}).

To reduce the order of the differential equations (\ref{emat}) by one,
we first split the index $a$ in such a way that $y^r$ is one of the
coordinates $y^a$, and $y^{\alpha}$ are the others. Then we impose the
conditions
\ba\label{cond} \p_{\alpha}\left(\frac{h_{\alpha\alpha}}{h_{aa}}\right)=0,\h
\p_{\alpha}\left(h_{rr}\dot{y}^r\right)^2 = 0,\h
\p_{r}\left(h_{\alpha\alpha}\dot{y}^{\alpha}\right)^2 = 0,\h
A^{\pm}_{\alpha}=\p_{\alpha}f^{\pm}.\ea
The result of integrations, compatible with (\ref{c2a}) and (\ref{mc}),
is the following
\ba\nl \left(h_{\alpha\alpha}\dot{y}^{\alpha}\right)^2 = D_{\alpha}
\left(y^a\ne y^{\alpha}\right) + h_{\alpha\alpha}\left[2\left(
A^{\pm}_{r}-\p_r f^{\pm}\right)\dot{y}^r - V\right]=E_{\alpha}
\left(y^{\beta}\right),\\ \nl
\left(h_{rr}\dot{z}^{r}\right)^2 = h_{rr}\left\{\left(\sum_{\alpha}-1\right)
V - \sum_{\alpha}\frac{D_{\alpha}}{h_{\alpha\alpha}}\right\} +
\left[\sum_{\alpha}\left(A^{\pm}_{r}-\p_r f^{\pm}\right)\right]^2 =
E_r\left(y^r\right),\ea
where $D_{\alpha}$, $E_{\alpha}$, $E_r$ are arbitrary functions of their
arguments, and
\ba\nl \dot{z}^r \equiv \dot{y}^r + \frac{\sum_{\alpha}}{h_{rr}}
\left(A^{\pm}_{r}-\p_r f^{\pm}\right).\ea
To find solutions of the above equations without choosing particular
metric, we have to fix all coordinates $y^a$ except one. If we denote it
by $y^A$, then the $exact$ solutions of the equations of motion and
constraints for a string in the considered curved background are given by
\ba\label{esec} X^{\mu}\left(X^A,\s\right)=X^{\mu}_{0} +
C^{\mu}_{\pm}\left(\lambda^1\tau+\sigma\right)-
\int_{X_0^A}^{X^A}k_0^{\mu\nu}\left[g^0_{\nu A}\mp A^{\pm}_{\nu}
\left(-\frac{h^0_{AA}}{V^0}\right)^{1/2}\right]d u ,\\ \nl
X^a=X^a_0=constants\h\mbox{for}\h a\ne A,\h
\tau\left(X^A\right)=\tau_0 \pm \int_{X_0^A}^{X^A}
\left(-\frac{h^0_{AA}}{V^0}\right)^{1/2}d u,\ea
where  $X^{\mu}_{0}$, ${X_0^A}$ and $\tau_0$ are arbitrary constants.
In these expressions
\ba\nl h^0_{AA}=h^0_{AA}\left(X^A\right)
=h_{AA}\left(X^A,X^{a\ne A}_0\right)\ea
and analogously for $V^0$, $k_0^{\mu\nu}$ and $g^0_{\nu A}$.

\section{Explicit example}
\hspace{1cm} In this section we give an explicit example of exact solution
for a string moving in four dimensional cosmological Kasner type background.
Namely, the line element is ($X^0 \equiv t$)
\ba\label{km} ds^2 &=& g_{MN}dX^M dX^N =
-(d t)^2 + \sum_{\mu=1}^{3}t^{2q_{\mu}}(dX^{\mu})^2,\\ \nl
&&\sum_{\mu=1}^{3}q_{\mu}=1,\h \sum_{\mu=1}^{3}q^2_{\mu}=1.\ea
For definiteness, we choose $q_{\mu}=\left(2/3,2/3,-1/3\right)$. The metric
(\ref{km}) depends on only one coordinate $t$, which we identify with
$X^a=y^a(\tau)$ according to our ansatz (\ref{ta}). Correspondingly, the
last two terms in (\ref{emat}) vanish and there is no need to impose the
conditions (\ref{cond}). Moreover, the metric (\ref{km}) is a diagonal one, so
we have $h_{aa}=g_{aa}=-1$. Taking this into account, we obtain the exact
solution of the equations of motion and constraints (\ref{esec}) in the
considered particular metric expressed as follows
\ba\nl X^{\mu}\left(t,\s\right)= X^{\mu}_0 +
C^{\mu}_{\pm}\left(\lambda^1\tau+\sigma\right)
\pm A^{\pm}_{\mu}I^{\mu}(t),\h \tau(t)=\tau_0 \pm I^0(t),\\ \nl
I^M (t)\equiv \int_{t_0}^{t}d u u^{-2q_M}V^{-1/2},\h
q_M = (0,2/3,2/3,-1/3).\ea
Although we have chosen relatively simple background metric, the expressions
for $I^M$ are too complicated. Because of that, we shall write down here
only the formulas for the two limiting cases $T=0$ and $T\to\infty$ for
$t\ge 0$. The former corresponds to considering null strings
(high energy string limit).

When $T=0$, $I^M$ reads
\ba\nl I^M = \frac{1}{2\left[\left(A_1^{\pm}\right)^2 +
\left(A_2^{\pm}\right)^2\right]^{1/2}}\Biggl[
\frac{t^{2/3-2q_M}}{\left(q_M-1/3\right)A}
\mbox{\scriptsize{2}}F_{1}\left(1/2,q_M-1/3;q_M+2/3;
-\frac{1}{A^2t^2}\right)\\ \nl + \frac{t_0^{5/3-2q_M}}{q_M-5/6}
\mbox{\scriptsize{2}}F_{1}\left(1/2,5/6-q_M;11/6-q_M;-A^2t_0^2\right) +
\frac{\Gamma\left(q_M-1/3\right)\Gamma\left(5/6-q_M\right)}
{\sqrt{\pi}A^{5/3-2q_M}}\Biggr],\ea where
\ba\nl A^2\equiv \frac{\left(A_3^{\pm}\right)^2}{\left(A_1^{\pm}\right)^2 +
\left(A_2^{\pm}\right)^2},\ea
$\mbox{\scriptsize{2}}F_{1}\left(a,b;c;z\right)$ is the Gauss'
hypergeometric function and $\Gamma(z)$ is the Euler's $\Gamma$-function.

When $T\to\infty$, $I^M$ is given by the equalities
\ba\nl I^{0} &=& \pm\frac{1}{4\lambda^0TC^3_{\pm}}\Biggl[\frac{6}{C}t^{1/3}
\mbox{\scriptsize{2}}F_{1}\left(1/2,-1/6;5/6;-\frac{1}{C^2t^2}\right)\\ \nl
&-&\frac{3}{2}t_0^{4/3}\mbox{\scriptsize{2}}F_{1}\left(1/2,2/3;5/3;
-C^2t_0^2\right) + \frac{\Gamma\left(-1/6\right)\Gamma\left(2/3\right)}
{\sqrt{\pi}C^{4/3}}\Biggr],\\ \nl
I^{1,2} &=& \pm\frac{1}{4\lambda^0TC^3_{\pm}}\left[\ln\left|
\frac{\left(1+C^2t^2\right)^{1/2}-1}{\left(1+C^2t^2\right)^{1/2}+1}\right| -
\ln\left|\frac{\left(1+C^2t_0^2\right)^{1/2}-1}
{\left(1+C^2t_0^2\right)^{1/2}+1}\right|\right],\\ \nl
I^3 &=& \pm\frac{1}{2\lambda^0TC^3_{\pm}C^2}\left[\left(1+C^2t^2\right)^{1/2}
- \left(1+C^2t_0^2\right)^{1/2}\right],\ea where
\ba\nl C^2\equiv \frac{\left(C^1_{\pm}\right)^2 + \left(C^2_{\pm}\right)^2}
{\left(C^3_{\pm}\right)^2}.\ea

\section{Comments and conclusions}
\hspace{1cm}In this letter we performed some investigation on the classical
string dynamics in $D$-dimensional curved background. In Section 2 we begin
with rewriting the string action in a form in which the limit $T\to 0$
could be taken to include also the null string case. Then we propose an
ansatz, which reduces the initial dynamical system depending on two
worldsheet parameters ($\tau,\sigma$) to the one depending only on $\tau$,
whenever the background metric does not depend at least on one coordinate.
An alternative ansatz is also given, which leads to a system depending only
on $\sigma$. In Section 3, using the existence of an abelian isometry group
$G$ generated by the Killing vectors $\p/\p X^{\mu}$, the problem of solving
the equations of motion and two constraints in $D$-dimensional curved
space-time $\mathcal{M}_D$ with metric $g_{MN}$ is reduced to considering
equations of motion and one constraint in the coset $\mathcal{M}_D/G$ with
metric $h_{ab}$. As might be expected, an interaction with a gauge field
appears in the Euler-Lagrange equations. In this connection, let us note
that if we write down $A^{\pm}_{a}$ introduced in (\ref{gp}) as
\ba\nl A^{\pm}_{a}=A_a^{\nu}A^{\pm}_{\nu},\ea
this gives the correspondence with the usual Kaluza-Klein type notation and
\ba\nl g_{MN}dy^M dy^N = h_{ab}dy^a dy^b + g_{\mu\nu}
\left(dy^\mu + A^\mu_a dy^a\right)\left(dy^\nu + A^\nu_b dy^b\right).\ea
In the remaining part of Section 3, we impose a number of conditions on the
background metric, sufficient to obtain exact solutions of the equations of
motion and constraints. These conditions are such that the metric
is general enough to include in itself many interesting cases
of curved backgrounds in different dimensions. In Section 4, we give an
explicit example of exact string solution in four dimensional Kasner type
background.


\end{document}